\documentclass{aa}
\usepackage{graphics}
\usepackage{times}
\newcommand{\rsun}{$\mathrm{R_{\sun}}$}
\newcommand{\msun}{$\mathrm{M_{\sun}}$}

\begin{document}

   \thesaurus{06 	    % A&A Section 6: Form. struct. and evolut. of stars
              (08.02.3;     % binaries: general,
               08.05.3;     % Stars: evolution, 
               08.13.2;     % Stars: mass-loss,
               03.13.4;     % Methods: numerical,
               08.16.7)}    % Pulsars: PSR J1454-5846.

   \title{On the energy equation and efficiency parameter of the 
          common envelope evolution}
%  \subtitle{}

   \author{Jasinta D. M. Dewi \inst{1,3,4}
      \and Thomas M. Tauris   \inst{2,1}}

   \offprints{jasinta@astro.uva.nl}

   \institute{Astronomical Institute {\it Anton Pannekoek},
              University of Amsterdam, Kruislaan 403, NL-1098 SJ Amsterdam,
              The Netherlands
         \and Nordic Institute for Theoretical Physics (NORDITA),
              Blegdamsvej 17, DK-2100 Copenhagen {\O}, Denmark
         \and Bosscha Observatory, Lembang 40391, Bandung Indonesia
         \and Department of Astronomy, Institut Teknologi Bandung, 
              Jl. Ganesha 10, Bandung 40132, Indonesia}

   \date{Received  / Accepted }

   \authorrunning{Dewi \& Tauris}
  
   \maketitle

%--------ABSTRACT--------------------------------------------------------------

\begin{abstract} 
We have investigated the structure of evolved giant stars with masses 3 -- 
10~\msun\ in order to evaluate the binding energy of the envelope to the core 
prior to mass transfer in close binary systems. This binding energy is expressed 
by a parameter $\lambda$ which is crucial for determining the outcome of 
binaries evolving through a common envelope (CE) and spiral-in phase. We discuss 
the $\lambda$--parameter and the efficiency of envelope ejection in the 
CE-phase, and show that $\lambda$ depends strongly on the evolutionary stage 
(i.e. stellar radius) of the donor star at the onset of the mass transfer. 
The existence of this relation enables us to introduce a new approach for 
solving the energy equation. For a given observed binary system we can derive a 
unique solution for the original mass and age of the donor star, as well as the 
pre-CE orbital period.\\
We find that the value of $\lambda$ is typically between 0.2 and 0.8. But in 
some cases, particularly on the asymptotic giant branch of lower-mass stars, it 
is possible that $\lambda > 5$. A high value of $\lambda$ (rather than {\it 
assuming} a high efficiency parameter, $\eta_\mathrm{CE} >1$) is sufficient to 
explain the long final orbital periods observed among those binary millisecond 
pulsars which are believed to have evolved through a CE-phase.\\ 
We also present a tabulation of $\lambda$ as a function of stellar radius and 
mass, which is useful for a quick estimation of the orbital decay during a 
common envelope and spiral-in phase. 

\keywords{stars: evolution -- stars: mass loss -- binaries: general 
          -- methods: numerical -- pulsars: PSR~J1454--5846} 
\end{abstract}

%          SECTION 1
%--------INTRODUCTION----------------------------------------------------------

\section{Introduction}
\label{lambda:sec:intro}

A very important stage of the evolution of close binaries is the formation of a 
common envelope. Generally, if a star in a binary fills its Roche-lobe (either 
because of the swelling up of the envelope due to exhaustion of nuclear fuel 
at the centre -- or reduction in the separation due to loss of orbital angular 
momentum) there will be an attempt to transfer mass to the companion. However, 
if the thermal re-adjustment timescale of the accreting star is larger than the 
mass-transfer timescale, the accreted layer piles up above the companion, heats 
up and expands so that the accreting star will also fill its Roche-lobe. The 
result is that the transfered matter will form a common envelope embedding both 
stars (Paczynski 1976; Ostriker 1976). When the mass ratio is large the CE-phase 
is accompanied by the creation of a drag-force, arising from the motion of the 
companion star through the envelope of the evolved star, which leads to 
dissipation of orbital angular momentum (spiral-in process) and deposition of 
orbital energy in the envelope. Hence, the global outcome of a CE-phase is 
reduction of the binary separation and often ejection of the envelope. For a 
general review on common envelopes, see e.g. Iben \& Livio (1993).

There is clear evidence of orbital shrinkage (as brought about by frictional 
torques in a CE-phase) in some observed close binary pulsars and white dwarf 
binaries (e.g. PSR 1913+16, L 870-2 -- see Iben \& Livio 1993 and references 
therein). In these systems it is clear that the precursor of the last-formed 
degenerate star must have achieved a radius much larger than the current orbital 
separation. Since the present short separation of degenerate stars in some 
binaries can not be explained by gravitational wave radiation or a magnetic wind 
as the main source of the loss of $J_\mathrm{orb}$, frictional angular momentum 
loss was probably responsible for the very close orbits observed. An alternative 
process recently investigated (King \& Begelman 1999; Tauris~et~al. 2000) is a 
highly Super-Eddington mass transfer on a subthermal timescale. However, as 
shown by the latter authors this process can not explain the observed properties 
of all systems. They conclude that CE and spiral-in evolution is still needed to 
account for the formation of double neutron star systems and mildly recycled 
pulsars with a heavy white dwarf companion and $P_\mathrm{orb} < 3$ days.

There are many uncertainties involved in calculations of the spiral-in phase 
during the CE evolution. The evolution is tidally unstable and the angular 
momentum transfer, dissipation of orbital energy and structural changes of the 
donor star take place on very short timescales ($10^3 - 10^4$ yr). A complete 
study of the problem would also require very detailed multi-dimensional 
hydrodynamical calculations which is beyond the scope of this paper. Here we aim 
to improve the simple formalism developed by Webbink (1984) and de Kool (1990) 
for estimating the orbital evolution of binaries evolving through a CE and 
spiral-in phase. An important quantity in their formalism is the binding energy 
between the envelope and the core of the donor star. This quantity is expressed 
by a parameter $\lambda$ which depends on the stellar density distribution, and 
consequently also on the evolutionary stage of the star. With a detailed 
evolutionary calculation, Bisscheroux (1998) found that $\lambda \sim 0.4 - 0.6$ 
on the red and asymptotic giant branches of a 5~\msun\ star. In the literature 
this parameter is usually taken to be a constant (e.g. $\lambda = 0.5$) 
disregarding the evolutionary stage (structure) of the donor star at the onset 
of the mass transfer process. This in turn leads to a misleading 
(over)estimation of the so-called efficiency parameter, $\eta_\mathrm{CE}$, 
often discussed by various authors. From observations one is able to put 
constraints on the product $\eta_\mathrm{CE} \lambda$ (see 
eq.~\ref{lambda:eq:webbink} below). However, by underestimating $\lambda$
the derived values for $\eta_\mathrm{CE}$ will be overestimated  -- e.g. 
van~den~Heuvel (1994) and Tauris (1996) concluded $\eta_\mathrm{CE} >1$ by 
assuming $\lambda = 0.5$. By calculating $\lambda$ as a real function of stellar 
radius we are able to improve the estimates of post-CE orbital separations and 
hence put more realistic constraints on $\eta_\mathrm{CE}$. 
 
Detailed calculations on the binding energy of the envelope to determine the 
outcome of binaries evolving through a CE-phase have been done by Han~et~al. 
(1994, 1995). The method and the stellar evolution code used to calculate the 
binding energy in their work is similar to the one applied in our work. However, 
here we discuss the binding energy within the context of the $\lambda$-parameter 
in the Webbink-formalism. We present calculated values of $\lambda$ for stars 
with initial mass in the range of 3 -- 10~\msun\ at different evolutionary 
stages. Such values are usefull for an estimation of the orbital decay during a 
common envelope and spiral-in phase. We also demonstrate a new approach to find 
a unique solution for the stellar and binary parameters of systems evolving 
through a CE and spiral-in phase.

We will in particular consider binaries in which the companion of the evolved 
star is a neutron star. Such binaries are the progenitors of binary millisecond 
pulsars (BMSPs) with a massive degenerate companion (e.g. van~den~Heuvel 1994) 
and they may appear in nature as short lived intermediate-mass {X}-ray binaries 
(IMXBs).

In Sect.~\ref{lambda:sec:ce-phase} we describe the basic concepts of common 
envelope evolution concerning the energy equation and the problems of defining 
the binding energy. Sect.~\ref{lambda:sec:code} introduces briefly the numerical 
stellar code used to evaluate the parameter $\lambda$. The results are presented 
in Sect.~\ref{lambda:sec:result} and discussed in 
Sect.~\ref{lambda:sec:discussion}. We summarize our conclusions in 
Sect.~\ref{lambda:sec:conclusion}.

%           SECTION 2
%--------COMMON ENVELOPE-------------------------------------------------------

\section{Common envelope and spiral-in phase}
\label{lambda:sec:ce-phase}

%        SUBSECTION 2.1
%--------ENERGY EQUATION-------------------------------------------------------

\subsection{The energy equation and orbital evolution}
\label{lambda:subsec:ce-equation}

A simple estimation on the reduction of the orbit can be found by simply 
equating the difference in orbital energy (before and after the CE-phase) to the 
binding energy of the envelope of the (sub)giant donor. Following the formalism 
of Webbink (1984), the original binding energy of the envelope at the onset of 
mass transfer will be of the order: $-G M_\mathrm{donor} M_\mathrm{env} / 
a_\mathrm{i} r_\mathrm{L}$, where $M_\mathrm{donor}$ is the mass of the donor 
star at the beginning of the CE-phase; $M_\mathrm{env}$ is the mass of the 
hydrogen-rich envelope of the donor; $a_\mathrm{i}$ is the orbital separation at 
the onset of the CE and $r_\mathrm{L} = R_\mathrm{L} / a_\mathrm{i}$ is the 
dimensionless Roche-lobe radius of the donor star, so that $a_\mathrm{i} 
r_\mathrm{L} = R_\mathrm{L} \approx R_\mathrm{donor}$. A parameter $\lambda$ was 
introduced by de Kool (1990) as a numerical factor (of order unity) which 
depends on the stellar density distribution, such that the real binding energy 
of the envelope can be expressed as:
	\begin{eqnarray}
	 E_\mathrm{env} & = & - 
	 \frac {G M_\mathrm{donor} M_\mathrm{env}} 
	       {\lambda a_\mathrm{i} r_\mathrm{L}} 
	 \label{lambda:eq:envkool}	 
	\end{eqnarray}
The total change in orbital energy is given by:
	\begin{eqnarray}
	 \Delta E_\mathrm{orb} & = & 
	 - \frac {G M_\mathrm{core} M_\mathrm{2}} {2 a_\mathrm{f}}  
	 + \frac {G M_\mathrm{donor} M_\mathrm{2}} {2 a_\mathrm{i}} 
	 \label{lambda:eq:orb}
	\end{eqnarray}
where $M_\mathrm{core} = M_\mathrm{donor} - M_\mathrm{env}$ is the mass of the 
helium core of the evolved donor star; $M_\mathrm{2}$ is the mass of the 
companion star and $a_\mathrm{f}$ is the final orbital separation after the 
CE-phase.

Let $\eta_\mathrm{CE}$ describe the efficiency of ejecting the envelope, i.e. 
of converting orbital energy into the kinetic energy that provides the outward 
motion of the envelope: $E_\mathrm{env} \equiv \eta_\mathrm{CE} \,\, \Delta 
E_\mathrm{orb}$ or, by equating eqs. (\ref{lambda:eq:envkool}) and 
(\ref{lambda:eq:orb}):
	\begin{eqnarray}
	 \frac {G M_\mathrm{donor} M_\mathrm{env}} 
	       {\lambda a_\mathrm{i} r_\mathrm{L}} & = &
	 \eta_\mathrm{CE} 
	 \left[ \frac {G M_\mathrm{core} M_\mathrm{2}} {2 a_\mathrm{f}} -
	 \frac {G M_\mathrm{donor} M_\mathrm{2}} {2 a_\mathrm{i}} \right]
	 \label{lambda:eq:orbind}
	\end{eqnarray}
which yields:
	\begin{eqnarray}
	  \frac {a_\mathrm{f}} {a_\mathrm{i}} & = & 
	  \frac {M_\mathrm{core} M_\mathrm{2}} {M_\mathrm{donor}}
	  \frac {1} {M_\mathrm{2} + 2 M_\mathrm{env}/
	  (\eta_\mathrm{CE}\lambda r_\mathrm{L})} 
	  \label{lambda:eq:webbink}
	\end{eqnarray}

%        SUBSECTION 2.2
%--------BINDING ENERGY--------------------------------------------------------

\subsection{The binding energy of the envelope}
\label{lambda:subsec:binding}

The total binding energy of the envelope to the core is given by:
	\begin{eqnarray}
	 E_\mathrm{bind} & = & 
	 \int_{M_\mathrm{core}}^{M_\mathrm{donor}} 
	 \left( - \frac {G M(r)} {r} + U \right) dm 
	 \label{lambda:eq:bind}
	\end{eqnarray}
where the first term is the gravitational binding energy and $U$ is the internal 
thermodynamic energy. The latter involves the basic thermal energy for a simple
perfect gas ($3 \Re T / 2 \mu$), the energy of radiation ($a T^{4} / 3 \rho$),
as well as terms due to ionization of atoms and dissociation of molecules
and the Fermi energy of a degenerate electron gas (Han~et~al. 1994, 1995).

The binding energy of the envelope can be expressed by eq. 
(\ref{lambda:eq:bind}), under the assumption that the entire internal energy is 
used efficiently in the ejection process. However, it is not known how much of
the internal energy is involved to unbind the envelope. Han~et~al. (1995) 
introduced a parameter $\alpha_\mathrm{th}$ and expressed the envelope binding 
energy as:
	\begin{eqnarray}
	 E_\mathrm{env} & = & - \int_{M_\mathrm{core}}^{M_\mathrm{donor}} 
	 \frac {G M(r)} {r} dm + \alpha_\mathrm{th} 		
	 \int_{M_\mathrm{core}}^{M_\mathrm{donor}} U dm 
	 \label{lambda:eq:envhan}
	\end{eqnarray}
The value of $\alpha_\mathrm{th}$ depends on the details of the ejection 
process, which is very uncertain. A value of $\alpha_\mathrm{th}$ equal to 0 or 
1 corresponds to maximum and minimum envelope binding energy, respectively. 
By simply equating eqs. (\ref{lambda:eq:envkool}) and (\ref{lambda:eq:envhan}) 
we calculated the parameter $\lambda$ for different evolutionary stages of a 
given star. The minimum and maximum derived values of $\lambda$ are denoted by 
$\lambda_\mathrm{g}$ ($\alpha_\mathrm{th}$ = 0) and $\lambda_\mathrm{b}$ 
($\alpha_\mathrm{th}$ = 1).

By taking the core mass $M_\mathrm{core}$ as the lower boundary of the integral, 
we assume that the structure of the core does not change during the envelope 
ejection process, and therefore we assume that there is no exchange of energy 
between the core and the envelope. Here, we define the core mass as the central 
mass which contains less than 10~\% hydrogen -- see discussion in 
Sect.~\ref{lambda:subsec:coremass}.

For the case of binary stars we use the radii of the stars to calculate the 
separations at the onset of the mass transfer assuming a companion star of mass 
$M_\mathrm{2}$. An estimation on the Roche-lobe radius is given by Eggleton 
(1983):
	\begin{eqnarray}
	 R_\mathrm{L} & = & 
	 \frac {0.49 \, a_\mathrm{i}} 
	       {0.6 + q^\mathrm{-2/3} \ln(1+q^\mathrm{1/3})} 
	 \label{lambda:eq:roche}
	\end{eqnarray}
where $q = M_\mathrm{donor} / M_\mathrm{2}$ is the mass ratio.

%        SECTION 3
%--------THE CODE--------------------------------------------------------------

\section{A brief introduction to the numerical computer code}
\label{lambda:sec:code}

We used an updated version of the numerical stellar evolution code of Eggleton 
(1971, 1972, 1973). This code uses a self-adaptive, non-Lagrangian mesh-spacing 
which is a function of local pressure, temperature, Lagrangian mass and radius. 
It treats both convective and semi-convective mixing as a diffusion process and 
finds a simultaneous and implicit solution of both the stellar structure 
equations and the diffusion equations for the chemical composition. New 
improvements are the inclusion of pressure ionization and Coulomb interactions 
in the equation-of-state, and the incorporation of recent opacity tables, 
nuclear reaction rates and neutrino loss rates. The most important recent 
updates of this code are described in Pols~et~al. (1995, 1998) and some are 
explained in Han~et~al. (1994).

We performed such detailed numerical stellar evolution calculations in our work
since they should yield more realistic results compared to models based on 
complete, composite or condensed polytropes.

%       SECTION 4
%--------RESULTS---------------------------------------------------------------

\setlength{\tabcolsep}{2.0pt}
	\begin{table}  
         \caption[]{The estimated $\lambda$ (see text) as a function of stellar 
          radius for different donor stars (graphs shown in 
          Fig.~\ref{lambda:fig:lambda}). $R$ is in units of \rsun\ and $k^2 =I / 
          M R^2$ is the gyration radius of the star. The corresponding orbital 
          period, $P_\mathrm{orb}$ (days) is also given at the onset of 
          Roche-lobe overflow, assuming a 1.3~\msun\ neutron star companion.}
	 \label{lambda:tab:periode}
	 \begin{center}
	 \begin{tabular}{rcrrrrcrcrrrr}
          \noalign{\smallskip}
	  \cline{1-6}\cline{8-13}
          \noalign{\smallskip}
	  \cline{1-6}\cline{8-13}
          \noalign{\smallskip}
	  \multicolumn{6}{c} {$M$ = 3.0 \msun} & &
	  \multicolumn{6}{c} {$M$ = 4.0 \msun} \\	 
          \noalign{\smallskip}
	  \cline{1-6}\cline{8-13}
          \noalign{\smallskip}
          $R$~~ & & $\lambda_\mathrm{g}$~ & $\lambda_\mathrm{b}$~ & $k^2$~ &
          $P_\mathrm{orb}$ & & $R$~~ & & $\lambda_\mathrm{g}$~ &
          $\lambda_\mathrm{b}$~ & $k^2$~ & $P_\mathrm{orb}$ \\
          \noalign{\smallskip}
	  \cline{1-6}\cline{8-13}
          \noalign{\smallskip}
          %
%           5.0 &         & 0.31 &  0.56 & 0.03 &   2.1 & & 
%                                6.0 &         & 0.32 &  0.58 & 0.03 &    2.2 \\
%          10.0 &         & 0.20 &  0.38 & 0.02 &   5.8 & & 
%                               10.0 &         & 0.22 &  0.42 & 0.02 &    4.8 \\
           20.0 & $^{*}$  & 0.46 &  0.87 & 0.13 &  16.4 & &
                                20.0 &         & 0.17 &  0.33 & 0.02 &   13.6 \\
           40.0 & $^{*}$  & 0.47 &  0.94 & 0.14 &  46.4 & & 
                                40.0 & $^{*}$  & 0.34 &  0.64 & 0.13 &   38.6 \\
           70.0 & $^{**}$ & 0.46 &  0.97 & 0.12 & 107.6 & & 
                                70.0 & $^{**}$ & 0.39 &  0.76 & 0.12 &   89.2 \\
          100.0 & $^{**}$ & 0.42 &  0.94 & 0.12 & 183.9 & & 
                               100.0 & $^{**}$ & 0.43 &  0.91 & 0.12 &  152.6 \\
          200.0 & $^{**}$ & 0.48 &  1.81 & 0.12 & 519.5 & & 
                               200.0 & $^{**}$ & 0.59 &  2.32 & 0.12 &  432.1 \\
          300.0 & $^{**}$ & 0.49 &  3.44 & 0.12 & 953.5 & & 
                               300.0 & $^{**}$ & 0.64 &  5.43 & 0.13 &  791.0 \\
          400.0 & $^{**}$ & 0.50 & 16.50 & 0.13 & 1470.0 & & 
                               400.0 & $^{**}$ & 0.59 & 10.60 & 0.14 & 1220.0 \\
          500.0 & $^{**}$ & 0.51 &       & 0.14 & 2048.9 & & 
                               500.0 & $^{**}$ & 0.61 &       & 0.15 & 1703.4 \\
          600.0 & $^{**}$ & 0.54 &       & 0.15 & 2700.7 & & 
                               600.0 & $^{**}$ & 0.63 &       & 0.16 & 2242.4 \\
          \noalign{\medskip}
          \noalign{\smallskip}
	  \cline{1-6}\cline{8-13}
          \noalign{\smallskip}
	  \cline{1-6}\cline{8-13}
          \noalign{\smallskip}
	  \multicolumn{6}{c} {$M$ = 5.0 \msun} & &
	  \multicolumn{6}{c} {$M$ = 6.0 \msun} \\	 
          \noalign{\smallskip}
	  \cline{1-6}\cline{8-13}
          \noalign{\smallskip}
          $R$~~ & & $\lambda_\mathrm{g}$~ & $\lambda_\mathrm{b}$~ & $k^2$~ &
          $P_\mathrm{orb}$ & & $R$~~ & & $\lambda_\mathrm{g}$~ &
          $\lambda_\mathrm{b}$~ & $k^2$~ & $P_\mathrm{orb}$ \\
          \noalign{\smallskip}
	  \cline{1-6}\cline{8-13}
          \noalign{\smallskip}
          %
%           6.5 &         & 0.34 &  0.60 & 0.03 &    2.2 & & 
%                                  7.0 &         & 0.35 & 0.62 & 0.03 &   2.2 \\
%          10.0 &         & 0.24 &  0.45 & 0.02 &    4.2 & & 
%                                 10.0 &         & 0.25 & 0.48 & 0.02 &   3.7 \\
           20.0 &         & 0.18 &  0.34 & 0.02 &   11.7 & & 
                                  20.0 &         & 0.18 & 0.36 & 0.02 &  10.4 \\
           40.0 &         & 0.22 &  0.41 & 0.04 &   33.3 & & 
                                  40.0 &         & 0.14 & 0.27 & 0.02 &  29.5 \\
           70.0 & $^{*}$  & 0.26 &  0.49 & 0.11 &   77.0 & & 
                                  70.0 & $^{*}$  & 0.19 & 0.35 & 0.11 &  68.3 \\
          100.0 & $^{*}$  & 0.28 &  0.54 & 0.13 &  131.0 & & 
                                 100.0 & $^{*}$  & 0.20 & 0.37 & 0.12 & 116.4 \\
          200.0 & $^{**}$ & 0.42 &  1.06 & 0.12 &  374.8 & & 
                                 200.0 & $^{**}$ & 0.27 & 0.55 & 0.11 & 332.3 \\
          300.0 & $^{**}$ & 0.67 &  4.03 & 0.13 &  687.7 & & 
                                 300.0 & $^{**}$ & 0.54 & 2.07 & 0.13 & 611.0 \\
          400.0 & $^{**}$ & 0.69 & 12.00 & 0.14 & 1058.0 & & 
                                 400.0 & $^{**}$ & 0.69 & 6.44 & 0.14 & 941.0 \\
          500.0 & $^{**}$ & 0.66 &       & 0.15 & 1490.0 & & 
                               500.0 & $^{**}$ & 0.71 & 44.87 & 0.15 & 1316.7 \\
          600.0 & $^{**}$ & 0.69 &       & 0.17 & 1960.5 & & 
                                600.0 & $^{**}$ & 0.70 &      & 0.18 & 1733.2 \\
          \noalign{\medskip}
          \noalign{\smallskip}
	  \cline{1-6}\cline{8-13}
          \noalign{\smallskip}
	  \cline{1-6}\cline{8-13}
          \noalign{\smallskip}
	  \multicolumn{6}{c} {$M$ = 7.0 \msun} & &
	  \multicolumn{6}{c} {$M$ = 8.0 \msun} \\	 
          \noalign{\smallskip}
	  \cline{1-6}\cline{8-13}
          \noalign{\smallskip}
          $R$~~ & & $\lambda_\mathrm{g}$~ & $\lambda_\mathrm{b}$~ & $k^2$~ &
          $P_\mathrm{orb}$ & & $R$~~ & & $\lambda_\mathrm{g}$~ &
          $\lambda_\mathrm{b}$~ & $k^2$~ & $P_\mathrm{orb}$ \\
          \noalign{\smallskip}
	  \cline{1-6}\cline{8-13}
          \noalign{\smallskip}
          %
%          10.0 &         & 0.27 & 0.51 & 0.03 &   3.3 & & 
%                                 10.0 &         & 0.28 & 0.54 & 0.02 &   3.0 \\
           20.0 &         & 0.19 & 0.37 & 0.02 &   9.4 & & 
                                  20.0 &         & 0.20 & 0.39 & 0.02 &   8.6 \\
           40.0 &         & 0.14 & 0.28 & 0.02 &  26.6 & & 
                                  40.0 &         & 0.14 & 0.29 & 0.02 &  24.4 \\
           70.0 &         & 0.11 & 0.22 & 0.02 &  61.6 & & 
                                  70.0 &         & 0.11 & 0.22 & 0.02 &  56.6 \\
          100.0 & $^{*}$  & 0.16 & 0.30 & 0.10 & 105.5 & & 
                                 100.0 &         & 0.10 & 0.19 & 0.02 &  96.2 \\
          200.0 & $^{**}$ & 0.15 & 0.28 & 0.11 & 300.7 & & 
                                 200.0 & $^{*}$  & 0.13 & 0.25 & 0.11 & 272.0 \\
          300.0 & $^{**}$ & 0.26 & 0.56 & 0.11 & 553.0 & & 
                                 300.0 & $^{**}$ & 0.17 & 0.32 & 0.11 & 508.5 \\
          400.0 & $^{**}$ & 0.65 & 3.93 & 0.13 & 853.0 & & 
                                 400.0 & $^{**}$ & 0.25 & 0.56 & 0.11 & 783.0 \\
          \noalign{\medskip}
          \noalign{\smallskip}
	  \cline{1-6}\cline{8-13}
          \noalign{\smallskip}
	  \cline{1-6}\cline{8-13}
          \noalign{\smallskip}
	  \multicolumn{6}{c} {$M$ =  9.0 \msun} & &
	  \multicolumn{6}{c} {$M$ = 10.0 \msun} \\	 
          \noalign{\smallskip}
	  \cline{1-6}\cline{8-13}
          \noalign{\smallskip}
          $R$~~ & & $\lambda_\mathrm{g}$~ & $\lambda_\mathrm{b}$~ & $k^2$~ &
          $P_\mathrm{orb}$ & & $R$~~ & & $\lambda_\mathrm{g}$~ &
          $\lambda_\mathrm{b}$~ & $k^2$~ & $P_\mathrm{orb}$ \\
          \noalign{\smallskip}
	  \cline{1-6}\cline{8-13}
          \noalign{\smallskip}
          %
%          10.0 &         & 0.29 & 0.56 & 0.03 &   2.8 & & 
%                                 10.0 &         & 0.34 & 0.61 & 0.03 &   2.6 \\
           20.0 &         & 0.20 & 0.40 & 0.02 &   8.0 & & 
                                  20.0 &         & 0.21 & 0.42 & 0.02 &   7.4 \\
           40.0 &         & 0.15 & 0.29 & 0.02 &  22.5 & & 
                                  40.0 &         & 0.15 & 0.31 & 0.02 &  21.1 \\
           70.0 &         & 0.11 & 0.23 & 0.02 &  52.5 & & 
                                  70.0 &         & 0.11 & 0.23 & 0.02 &  48.7 \\
          100.0 &         & 0.09 & 0.19 & 0.02 &  89.4 & & 
                                 100.0 &         & 0.10 & 0.20 & 0.02 &  83.8 \\
          200.0 & $^{*}$  & 0.11 & 0.21 & 0.10 & 252.3 & & 
                                 200.0 & $^{*}$  & 0.11 & 0.22 & 0.08 & 235.2 \\
          300.0  &$^{*}$  & 0.13 & 0.25 & 0.12 & 465.7 & & 
                                 300.0 & $^{*}$  & 0.10 & 0.19 & 0.10 & 432.2 \\
          400.0 & $^{**}$ & 0.17 & 0.33 & 0.11 & 727.9 & & 
                                 400.0 & $^{**}$ & 0.11 & 0.20 & 0.10 & 681.0 \\
          500.0 & $^{**}$ & 0.25 & 0.59 & 0.12 & 1013.3 & & 
                                 500.0 & $^{**}$ & 0.17 & 0.35 & 0.11 & 952.6 \\
          \noalign{\medskip}
	  \cline{1-6}\cline{8-13}
          \noalign{\smallskip}
	  \cline{1-6}\cline{8-13}
         \end{tabular}
	 \end{center}
	 \begin{list}{}{}
	 \item[$^{*}$]  Red giant branch
	 \item[$^{**}$] Asymptotic giant branch
	 \end{list}
	\end{table}
\setlength{\tabcolsep}{6pt}

\section{Results}
\label{lambda:sec:result}

The above-mentioned detailed stellar evolution code has been used to probe the 
structure of donor stars at the onset of mass transfer. We evolved stars with an 
initial mass in the interval of 3 -- 10~\msun\ assuming a chemical composition 
of (${X} = 0.70, {Z} = 0.02$) and using a mixing-length parameter of $\alpha = 
l/H_{p} = 2.0$. Convective overshooting is taken into account in the same way as 
in Pols~et~al. (1998) using an overshooting constant $\delta_\mathrm{ov} = 0.1$. 
We used de~Jager's wind mass-loss rate (de~Jager~et~al. 1988; Nieuwenhuijzen \& 
de~Jager 1990) to estimate the mass loss prior to the Roche-lobe overflow and CE 
evolution. The wind mass loss from the donor star (and subsequent widening of 
the orbit) prior to the Roche-lobe overflow is typically only $\sim 1 - 4$~\% on 
the AGB and hence this effect is not very significant for the stars considered 
here.

	\begin{figure*}
 	 \centerline{\resizebox{16cm}{!}{\includegraphics{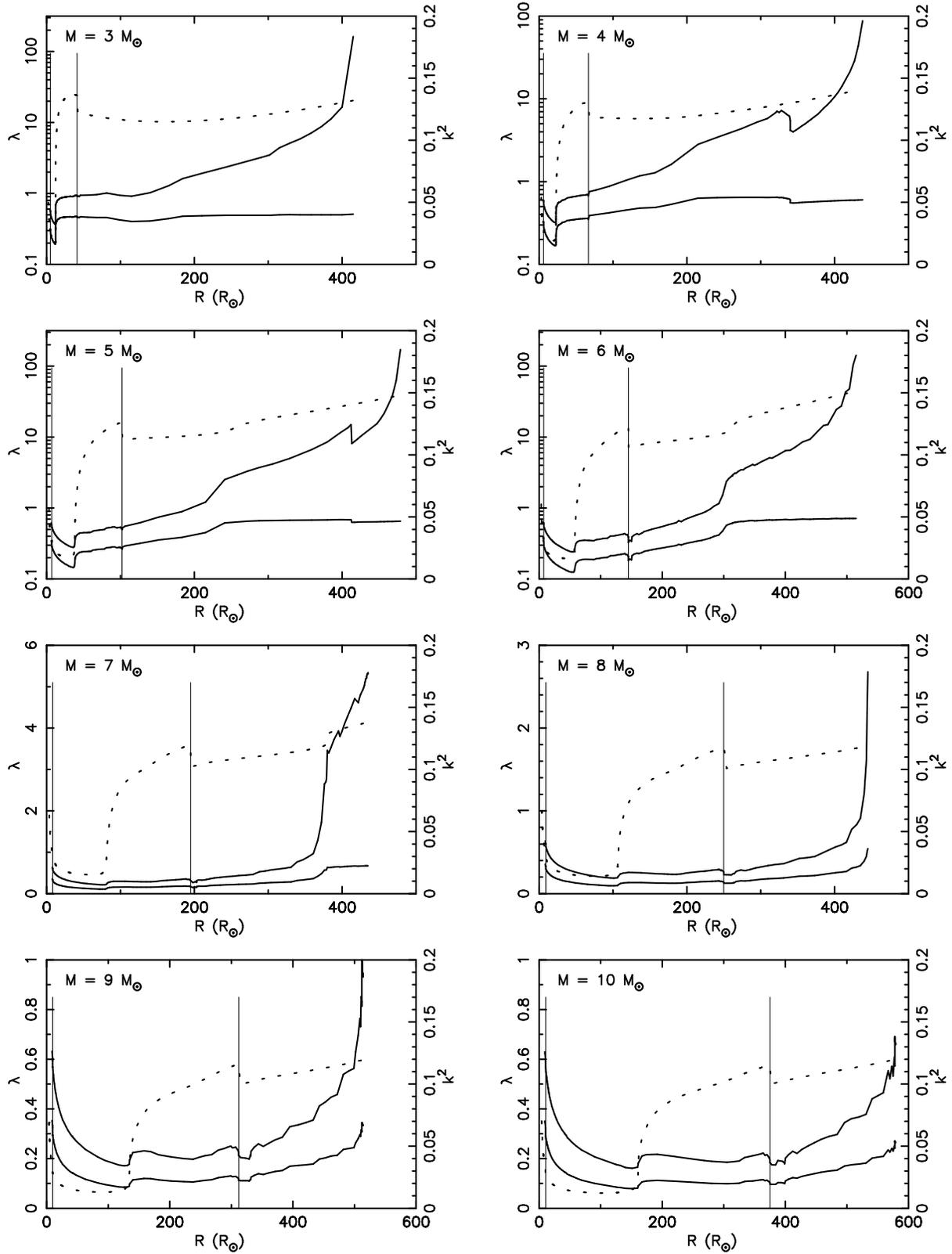}}}
	 \caption{The parameter $\lambda$ as a function of stellar radius, for
	  stars with initial mass of 3 -- 10~\msun. Upper and lower solid lines
	  represent the value of $\lambda$ derived from the total
	  ($\lambda_\mathrm{b}$) and gravitational ($\lambda_\mathrm{g}$)
	  binding energy, respectively -- see text. For stars with mass in the
	  interval of 3 -- 6~\msun, the $\lambda_\mathrm{b}$-curves are plotted
	  until the moment before the binding energy becomes positive (and the 
	  $\lambda_\mathrm{b}$-value would become negative, see 
          Sect.~\ref{lambda:subsec:question} for an explanation). The dotted 
          line in each panel presents the gyration radius of the star 
          ($k^2=I/MR^2$) with a scale given on the right y-axis. The vertical 
          lines separate the regions of case A (left-), case B (middle-) and 
          case C (right-side) mass transfer from a binary donor.}
	 \label{lambda:fig:lambda}
	\end{figure*}

	\begin{figure*}
 	 \centerline{\resizebox{14cm}{!}{\includegraphics{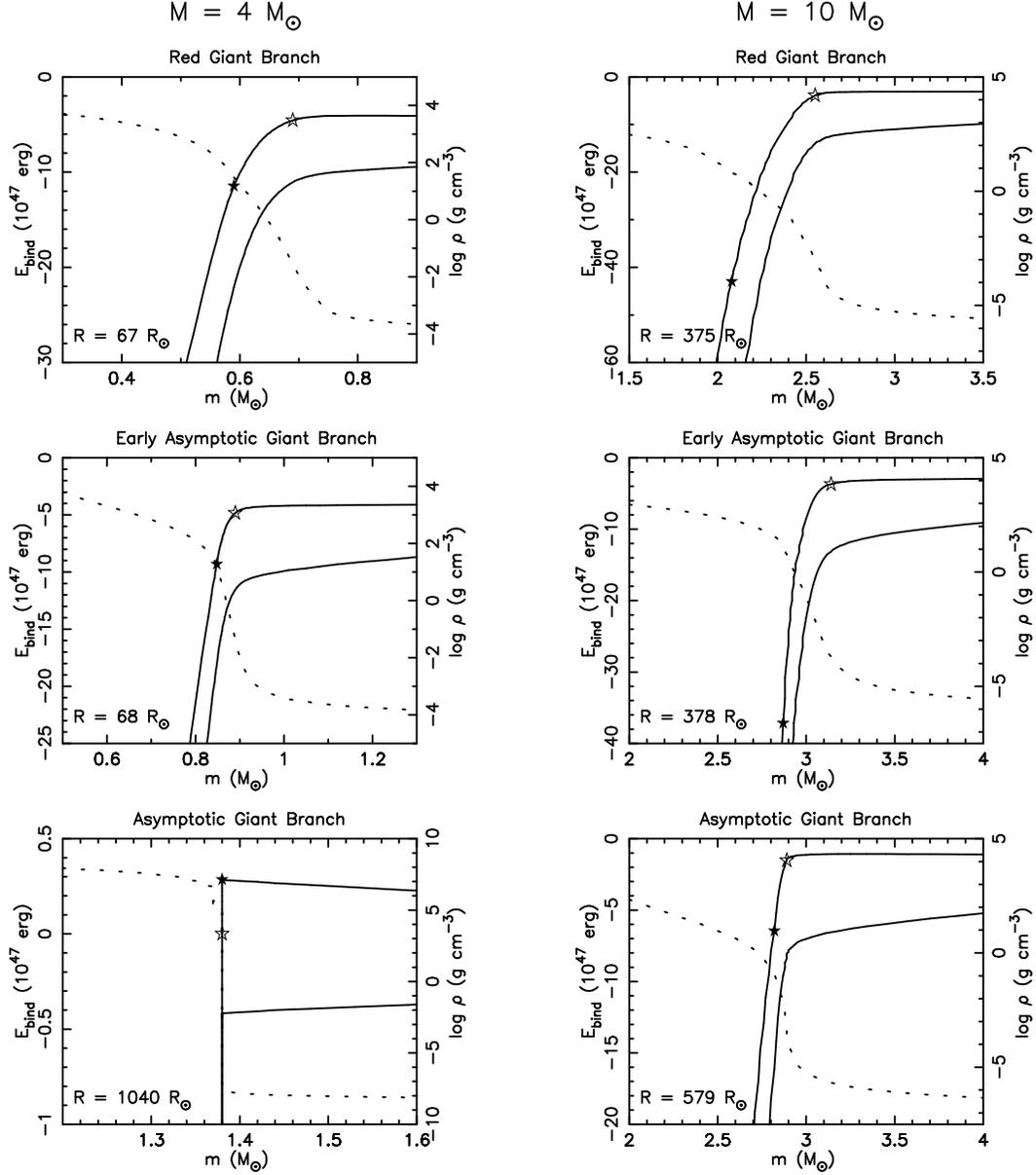}}}
	 \caption{Total and gravitational binding energy (upper and lower solid
	  line, respectively) around the core of stars with an initial mass
	  of 4~\msun\ (left-side panels) and 10~\msun\ (right-side panels) at
	  three giant stages. The stellar radius corresponding to each stage is
	  indicated at the bottom of the panels. The density profile is
	  represented by a dotted line (with the scale given on the right
	  y-axis). Solid and open stars indicate the core mass chosen to 
          determine the binding energy using our choice and the definition of 
          Han~et~al. (1994), respectively.}
	 \label{lambda:fig:density}
	\end{figure*}

With the Eggleton code, the total binding energy is computed for each 
mesh-point. Included here are the ionization of $\mathrm{H^{+}}$, H, 
$\mathrm{He^{++}}$, $\mathrm{He^{+}}$, He, dissociation of $\mathrm{H_{2}}$, 
rotational and vibrational modes of $\mathrm{H_{2}}$, as well as ionization of 
seven heavier elements (C, N, O, Ne, Mg, Si, Fe) which are assumed to be fully 
ionized at all temperatures and densities (Pols~et~al. 1995). By integrating 
from the core to the photosphere, we found the total binding energy of the 
envelope. We used this energy to derive the parameter $\lambda$. In 
Fig.~\ref{lambda:fig:lambda} we present the relation between $\lambda$ and 
stellar radius for each of the different donor stars considered. Here $\lambda$ 
was determined either from the total binding energy of the envelope (i.e. 
$\alpha_\mathrm{th} = 1$) or from the gravitational binding energy alone (i.e. 
$\alpha_\mathrm{th} = 0$).

For a donor star in a binary system the stellar radius at the onset of the 
CE-phase is roughly equivalent to its Roche-lobe radius, and given a mass of its 
companion star one finds the corresponding orbital period of the system. In 
Table~\ref{lambda:tab:periode}, assuming that the companion is a 1.3~\msun\ 
neutron star, we list the relation between the orbital period of the binary and 
$\lambda$ at the onset of the mass transfer (again due to the wind mass loss, 
this period might be slightly larger than the initial ZAMS period). We 
distinguish $\lambda_\mathrm{g}$ (as derived from gravitational binding energy 
alone) from $\lambda_\mathrm{b}$ (calculated from the total binding energy). The 
gyration radius of the star, $k^2 = I / M R^2$, is also presented to provide 
information about the orbital tidal instability. Mass transfer becomes tidally 
unstable if the rotational angular momentum of the donor exceeds one third of 
the orbital angular momentum (Darwin 1908).

%        SECTION 5
%-------DISCUSSION-------------------------------------------------------------

\section{Discussion}
\label{lambda:sec:discussion}

%       SUBSECTION 5.1
%-------BINDING ENERGY---------------------------------------------------------

\subsection{On the binding energy of the envelope}
\label{lambda:subsec:question}

In Fig.~\ref{lambda:fig:density} we have plotted the binding energy of the 
envelope as well as the density profile of two stars with an initial mass of 4 
and 10~\msun. When the star reaches the asymptotic giant branch, the core is 
surrounded by a convective envelope, and the density gradient around the core 
mass is very steep. As the star ascends this giant branch, the convective
envelope becomes deeper and the density contrast higher. 

Assuming a star to be a non-rotating globe of non-degenerate ionized hydrogen 
and helium in hydrostatic equilibrium (i.e. nearly an ideal monatomic gas) one 
would expect from the virial theorem: $\lambda_\mathrm{b} \simeq 2 
\lambda_\mathrm{g}$. However at late stellar evolutionary stages this picture no 
longer holds. After the second dredge-up stage of evolution, the gravitational 
binding energy becomes smaller as the stellar radius expands to a supergiant 
dimension. Meanwhile the ionization and dissociation energy becomes more 
significant when the temperature decreases in the envelope (Bisscheroux 1998) -- 
raising the internal energy above 0.5 times the gravitational binding energy in 
absolute value. Hence, the total binding energy becomes smaller in absolute 
value relative to the gravitational binding energy. This in turn causes a larger 
discrepancy between $\lambda_\mathrm{g}$ and $\lambda_\mathrm{b}$, see upper 
right panel in Fig.~\ref{lambda:fig:lambda}.

Especially for low-mass stars (3 -- 6~\msun) we see that the contribution
of the internal energy to the total binding energy becomes dominating. 
At some point on the asymptotic giant branch, the internal energy dominates over 
the gravitational binding energy, and hence the total binding energy of the 
envelope reaches a positive value. One might regard this alteration from a 
negative to a positive value of the total binding energy as the ejection of the 
giant's envelope, or as a point of the onset of a pulsationally driven (Mira) 
superwind (Han~et~al. 1994 and references therein). However, when the binding 
energy becomes positive, the derived parameter $\lambda_\mathrm{b}$ will be
negative and our formalism in Sect.~\ref{lambda:subsec:ce-equation} breaks down 
-- cf. resulting ''negative'' final separation in eq.~(\ref{lambda:eq:webbink}). 
On the contrary $\lambda_\mathrm{g}$ is relatively constant during this late 
stage of evolution.

The evolution of stars above 7~\msun\ is cut short by carbon burning. Therefore 
they never move as far up the AGB as lower mass stars, and the internal 
(recombination) energy never becomes so important as to dominate the total 
binding energy. This is why the difference between $\lambda_\mathrm{g}$ and 
$\lambda_\mathrm{b}$ is less pronounced as we go to the more massive stars.

%     SUBSECTION 5.2
%-------CORE MASS---------------------------------------------------------------

\subsection{On the definition of the core mass boundary}
\label{lambda:subsec:coremass}

Our results are sensitive to the definition of core mass. In our work we have 
chosen to define the core mass as the central mass which contains less than 
10~\% hydrogen. Using other boundary conditions such as $\partial ^2 \log \rho / 
\partial m^2 = 0$ (e.g. Bisscheroux 1998) or the alternative method by 
Han~et~al. (1994) will lead to different values for the binding energy of the 
envelope and hence different estimates of $\lambda$. In general the above 
mentioned methods will result in a core mass boundary further out in the star 
compared to our definition (see symbols on the dashed line in 
Fig.~\ref{lambda:fig:density}). Hence their methods lead to smaller envelope 
binding energies and thus larger values of $\lambda$ (factor $\sim 2$) on the 
RGB and EAGB. For a 10~\msun\ star, this difference is much higher than a factor 
of 2. Further up the AGB, where the density gradient is steeper, the differences 
become smaller and the result more reliable. Nevertheless, in accordance with 
our findings using any of the two other boundary conditions mentioned will also 
result in large values of $\lambda$ on the AGB (for low-mass stars: 
$\lambda_\mathrm{b} \gg 1$) and this is the important issue. 

%              SUBSECTION 5.3
%---------CHEMICAL AND MIXING-LENGTH--------------------------------------------

\subsection{Dependency of the $\lambda$-parameter on the chemical composition 
            and mixing-length parameter}
\label{lambda:subsec:composition}

\begin{figure}
 	 \centerline{\resizebox{7.5cm}{!}{\includegraphics{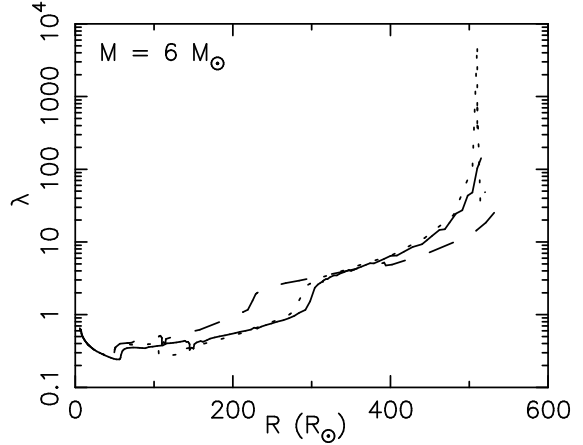}}}
	 \caption[]{$\lambda_\mathrm{b}$ as a function of radius for a 6~\msun\ 
	  star. The full line is for ${X} = 0.70, {Z} = 0.02$ and $\alpha = 
	  2.0$. The dashed line is for the same chemical composition with 
	  $\alpha = 3.0$. The dotted line is for a star with ${X} = 0.75, 
	  {Z} = 0.001$ and $\alpha =2.0$. See text for a discussion.}
	 \label{lambda:fig:chemical}
	\end{figure}

In order to investigate the dependency of the $\lambda$-parameter on chemical 
composition and mixing-length parameter, we have plotted in 
Fig.~\ref{lambda:fig:chemical} the computed $\lambda_\mathrm{b} (R)$ for both 
Pop.~{I} and Pop.~{II} chemical abundancies, as well as for two different values 
of the mixing-length parameter for a 6~\msun\ donor star.

>From the ZAMS to the base of the RGB ($R \sim 50$~\msun) there are no 
differences on the calculated values of $\lambda_\mathrm{b}$. It is also seen 
that there is hardly any dependency on the chemical composition for the rest of 
the evolutionary track far up the AGB. However, from the horizontal branch ($R 
\sim 150$~\msun) and halfway up the AGB ($R \sim 300$~\msun) a higher value of 
$\lambda_\mathrm{b}$ is found using a larger value of the mixing-length 
parameter. This picture is reversed near the top of the AGB ($R > 400$~\msun). 
In general we can conclude that $\lambda_\mathrm{b}$ is almost independent of 
the chemical composition of the star, but it depends on the choice of the 
mixing-length parameter.

%           SUBSECTION 5.4
%--------EFFICIENCY PARAMETER---------------------------------------------------

\subsection{The efficiency parameter of the ejection process}
\label{lambda:subsec:efficiency}

	\begin{figure}
 	 \centerline{\resizebox{7.5cm}{!}{\includegraphics{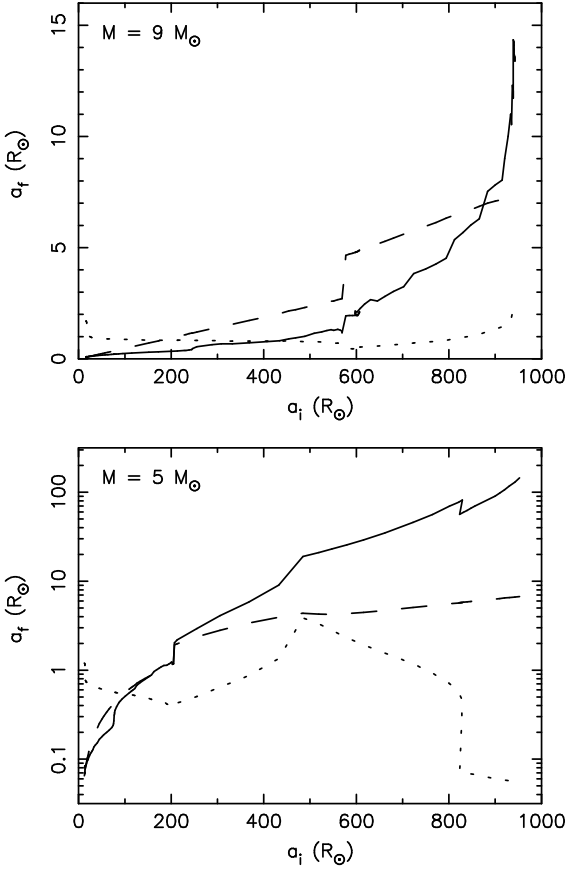}}}
	 \caption[]{Final orbital separations after a CE-phase for systems with
          an initial donor star of mass 9~\msun\ and 5~\msun\ (upper and lower
          panel, respectively). The companion is assumed to be a 1.3~\msun\
          neutron star. The full line is the final separation we obtain when the
          parameter $\lambda$ was calculated as a function of stellar radius and
          the dashed line is for a constant $\lambda = 0.5$. The dotted line is
          the critical final separation at which the helium star (the core of 
          the stripped donor) immediately fills its Roche-lobe after the CE and
          spiral-in phase. If the final separation, $a_\mathrm{f}$ is smaller
          than this value the system is most likely to merge and hence the 
          binary will not survive. The radius of the helium star was calculated 
          with our numerical stellar evolution code.}
	 \label{lambda:fig:separation}
	\end{figure}
	
In order to explain the long orbital periods observed among those of the BMSPs 
which are likely to have evolved through a CE-phase, Tauris (1996) and 
van~den~Heuvel (1994) found it necessary to invoke efficiency parameters larger 
than unity (e.g. $1 < \eta_\mathrm{CE} < 4$). However, their simple estimates 
were based on a constant value of $\lambda = 0.5$. Here we suggest that a high 
value of $\lambda$ (rather than a high value of $\eta_\mathrm{CE}$) is the 
natural explanation for the long orbital periods observed. This in turn 
indicates that if the internal energy is used efficiently ($\alpha_\mathrm{th} = 
1$) to unbind the envelope, there is no need to assume $\eta_\mathrm{CE} > 1$ 
for these systems. This result was also concluded by Han~et~al. (1995).

Let us now consider a system with a 9~\msun\ donor star and a 1.3~\msun\ neutron 
star. Such a binary is the progenitor of a pulsar with a massive degenerate 
companion. We computed the final orbital separation of the binary after a 
CE-phase by using $E_\mathrm{env} = \eta_\mathrm{CE} \,\, \Delta E_\mathrm{orb}$ 
combined with eqs. (\ref{lambda:eq:orb}) and (\ref{lambda:eq:envhan}). Here we 
applied $\eta_\mathrm{CE} = 1$ and $\alpha_\mathrm{th} = 1$ (i.e. including the
internal thermodynamic energy, $\lambda=\lambda_\mathrm{b}$), and present the
result in the upper panel of Fig.~\ref{lambda:fig:separation}. As a comparison, 
we also show the final separation computed by using 
eq.~(\ref{lambda:eq:webbink}) with a constant efficiency parameter $\lambda = 
0.5$ (commonly used in the literature).

A binary will survive the CE-phase only if the final orbit is large enough so 
that the core of the donor will not instantly fill its Roche-lobe. In the case 
of a binary with a 9~\msun\ donor and a 1.3~\msun\ neutron star, this condition 
is reached if the binary separation at the onset of the CE-phase is larger than 
435~\rsun\ ($P_\mathrm{orb} > 327^\mathrm{d}$). If we use the constant $\lambda 
= 0.5$, this separation is reduced to 175~\rsun\ ($P_\mathrm{orb} > 
84^\mathrm{d}$). In the latter case the donor star still has a radiative 
envelope with only a very thin convective shell at the onset of the mass 
transfer. Therefore the CE is formed only as a result of orbital shrinkage due 
to mass transfer with a high mass ratio between the donor and the neutron star. 
In the former case, the donor star had time to develop a deep convective 
envelope and it will therefore expand in response to mass loss due to its 
isentropic entropy profile. This expansion, in combination with the orbital 
shrinkage mentioned above, will immediately lead to a runaway mass transfer, and 
hence a CE will be formed easily.\\
At late stages of stellar evolution, the value of $\lambda_\mathrm{b}$ has an 
important effect on the final orbital period. This effect is more significant in 
binaries with less massive donors as can be seen in the lower panel of 
Fig.~\ref{lambda:fig:separation} for the case of a 5~\msun\ donor. 

%      SUBSECTION 5.5
%--------NEW METHOD-------------------------------------------------------------

\subsection{A new approach for tracking the evolutionary history of a system
            evolving through a CE}
\label{lambda:subsec:newmethod}

As we have demonstrated in this paper, $\lambda$ is a function of stellar radius
for any given mass. Combining this fact with eq.~(\ref{lambda:eq:webbink}),
which relates the observed orbital period to the original radius of the
Roche-lobe filling donor star as a function of mass and $\lambda$ (or 
equivalently: $a_\mathrm{f}/a_\mathrm{i}$), we are able to place severe 
constraints on the mass and radius (and hence age) of the original donor. 

	\begin{figure*}
	\begin{center}
 	 \centerline{\resizebox{10cm}{!}{\includegraphics{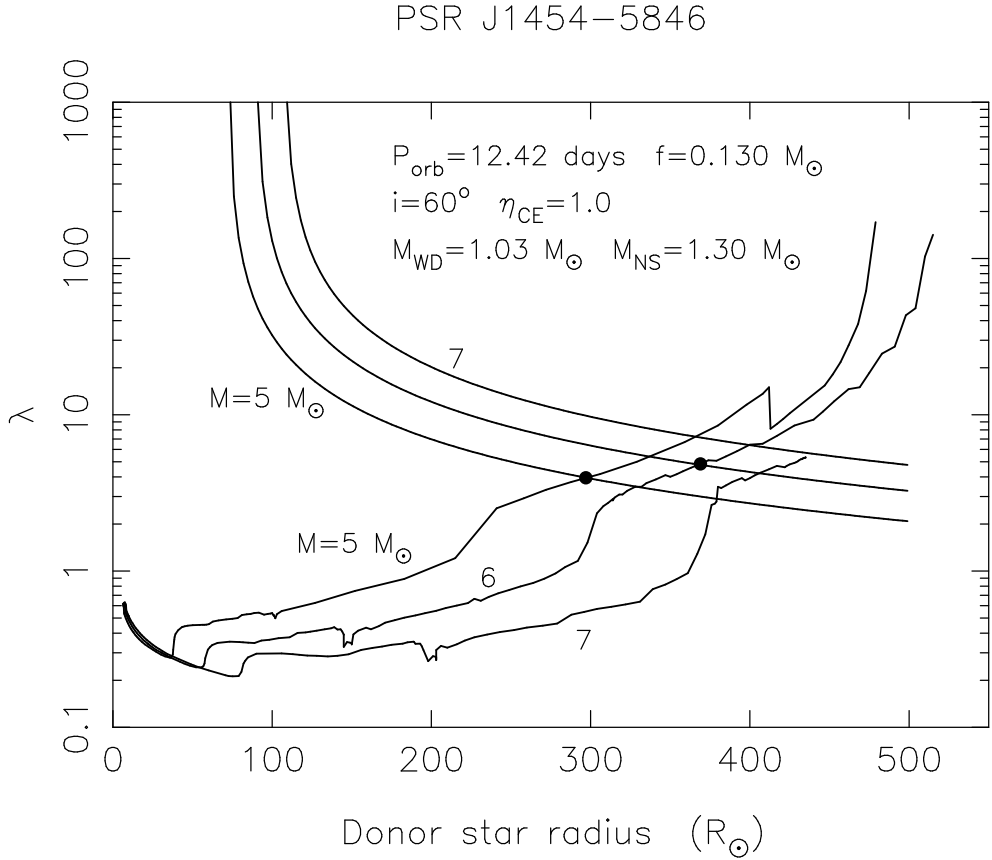}}}
	 \caption[]{The parameter $\lambda = \lambda_\mathrm{b}$ as a function 
          of stellar radius, for donor stars with initial mass of 5 -- 7~\msun.
          Also shown are curves for eq.~(\ref{lambda:eq:webbink}) calculated 
          with the observed parameters given for PSR~J1454--5846 and assuming
          $\eta_\mathrm{CE} = 1.0$. The intersection of these curves yields a 
          unique solution for the properties of the original donor star (the 
          progenitor of the white dwarf) and the pre-CE orbital period -- see 
          text for a discussion.}
	 \label{lambda:fig:newmethod}
	 \end{center}
	\end{figure*}
	
An example of this approach is shown in Fig.~\ref{lambda:fig:newmethod} where we 
have used the parameters for the recently discovered binary pulsar 
PSR~J1454--5846 (Camilo~et~al. 2000). This system has an orbital period of 
$P_\mathrm{orb} = 12 \fd 42$ and a mass function, $f = 0.130$~\msun. The exact 
mass of the white dwarf in this system is unknown; but as an example we shall 
assume an orbital inclination angle, $i = 60\degr$ yielding $M_\mathrm{WD} = 
1.03$~\msun (for $M_\mathrm{NS} = 1.30$~\msun). Hence we must require that the 
original giant donor star had a core mass of this value by the time its envelope 
was ejected in the CE-phase. In Fig.~\ref{lambda:fig:newmethod} the intersection 
points for a 5~\msun\ and 6~\msun\ donor yield core mass values of 0.93~\msun\ 
and 1.10~\msun, respectively. To obtain a core mass of 1.03~\msun\ we must 
therefore search for a solution in-between. We find $M_\mathrm{donor} = 
5.6$~\msun\ and the radius at the onset of the RLO is 340~\rsun. The 
corresponding age of the donor is 93~Myr and the pre-CE orbital period is 
760$^\mathrm{d}$.

\begin{figure}
 	 \centerline{\resizebox{6.5cm}{!}{\includegraphics{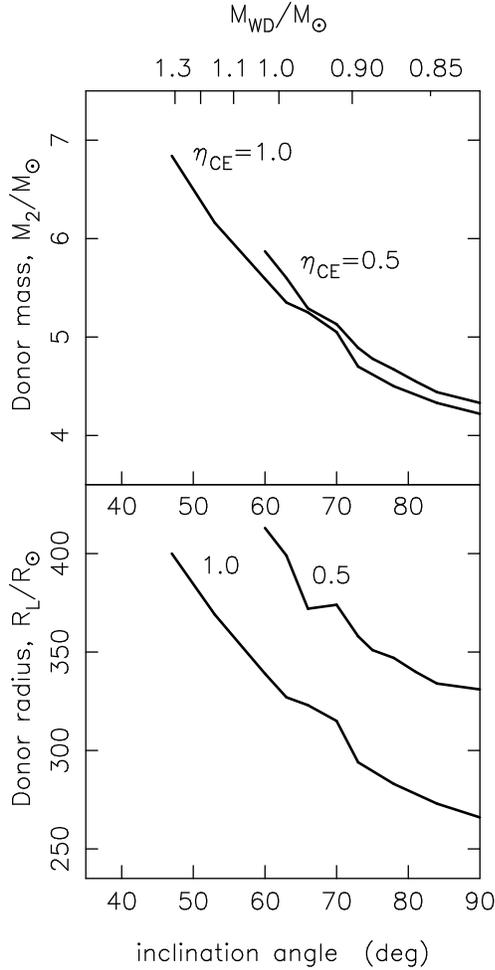}}}
	 \caption[]{Solutions for the mass (top panel) and radius (bottom panel)
          of the giant progenitor of the massive white dwarf observed in 
          PSR~J1454--5846, as a function of orbital inclination angle, $i$ and 
          CE efficiency parameter, $\eta_\mathrm{CE}$. On the top is shown 
          $M_\mathrm{WD}$ as a function of the inclination angle -- see text for 
          further details.}
	 \label{lambda:fig:1454-5846}
	\end{figure}

The above solution was found for an inclination angle, $i = 60\degr$. In 
Fig.~\ref{lambda:fig:1454-5846} we have plotted the solutions for 
$M_\mathrm{donor}$ and $R_\mathrm{L}$ as a function of inclination angle for two 
different values of the efficiency parameter, $\eta_\mathrm{CE}$. It is 
interesting to notice that $M_\mathrm{donor}$ is more or less independent of 
$\eta_\mathrm{CE}$ whereas $R_\mathrm{L}$ gets larger for smaller values of 
$\eta_\mathrm{CE}$. This result is expected theoretically since a low value of 
$\eta_\mathrm{CE}$ requires either more orbital energy released and/or less 
binding energy of the envelope, and thus a larger $R_\mathrm{L}$, in order to 
eject the envelope successfully. An upper limit to the mass of the white dwarf 
is about 1.3~\msun\ which yields $i > 47\degr$. For $\eta_\mathrm{CE} = 0.5$ no 
solution is found for $i < 59\degr$. Hence it would be interesting if a future 
detection of a spectroscopic line from the white dwarf would yield an orbital 
inclination angle less than this value, since small values of $\eta_\mathrm{CE}$ 
would be excluded that way.

It should also be noted from Fig.~\ref{lambda:fig:1454-5846} that the binary 
pulsar must have had an AGB radius at the onset of the RLO. Many of the observed 
binary pulsars which have survived a common envelope evolution also have long 
orbital periods. Hence, the density gradient was fairly steep at the onset of 
the Roche-lobe overflow and therefore the uncertainties regarding the core mass 
definition as already discussed in Sect.~\ref{lambda:subsec:coremass} are less 
important in this context.

%        SECTION 6
%-------CONCLUSION-------------------------------------------------------------

\section{Conclusions}
\label{lambda:sec:conclusion}

We have adapted a numerical computer code to study the stellar structure of 
stars with an initial mass of 3 -- 10~\msun\ in order to evaluate the binding 
energy of the envelope to the core which determines the parameter $\lambda$ in 
the energy equation of the common envelope evolution. We have presented evidence 
that the parameter $\lambda$ depends strongly on the evolutionary stage. Hence 
the value of $\lambda$, at the onset of the mass transfer and CE-phase, is 
unique for binaries with different initial orbital periods and donor masses. 
Taking advantage of this, we have demonstrated a new approach for finding a 
unique solution to the energy equation and the original mass and radius of the 
donor star. An application of this approach on PSR~J1454--5846 yields a 
constraint for the donor star and the pre-CE orbital period as a function of the 
unknown orbital inclination angle.

We have demonstrated that in order to obtain a long final orbital period after 
a CE-phase (as observed in some BMSPs), we do not require an efficiency 
parameter larger than unity, if the internal energy can be used efficiently to 
eject the envelope.

In order to obtain the final orbital separation of a binary system after a CE 
and spiral-in phase, we advise one to calculate the binding energy from the 
stellar structure by means of eq. (\ref{lambda:eq:envhan}) with $0 \leq 
\alpha_\mathrm{th} \leq 1$. In case the detailed structure of the donor is not 
available (e.g. for a quick back-of-the-envelope-calculation), 
eq.~(\ref{lambda:eq:webbink}) can be used with $\eta_\mathrm{CE} = 1$ in 
combination with $\lambda_\mathrm{g} \leq \lambda \leq \lambda_\mathrm{b}$. The 
value of $\lambda$ can be found in our Table \ref{lambda:tab:periode} for a 
given donor star mass and radius.

%--------ACKNOWLEDGEMENT & REFERENCES------------------------------------------

\begin{acknowledgements}
This work was sponsored by NWO Spinoza Grant 08-0 to E.~P.~J. van~den~Heuvel. 
The authors thank E.~P.~J. van~den~Heuvel, Gertjan Savonije, Gijs Nelemans and 
Winardi Sutantyo for stimulating and fruitful discussions; and Onno Pols for 
his valuable referee suggestions and comments. T.~M.~T. acknowledges the receipt 
of a NORDITA fellowship.
\end{acknowledgements}


\begin{thebibliography}{}
  \bibitem{} Bisscheroux B., 1998, M.Sc. Thesis, Univ. Amsterdam
  \bibitem{} Camilo F., et al., 2000 in preparation
  \bibitem{} Darwin G. H., 1908, {\it Scientific Papers}, vol. 2, 
             Cambridge Univ. Press
  \bibitem{} de Jager C., Nieuwenhuijzen H., van der Hucht K. A., 1988, 
             A\&AS 72, 259
  \bibitem{} de Kool M., 1990, ApJ 358, 189
  \bibitem{} Eggleton P. P., 1971, MNRAS 151, 351
  \bibitem{} Eggleton P. P., 1972, MNRAS 156, 361
  \bibitem{} Eggleton P. P., 1973, MNRAS 163, 279
  \bibitem{} Eggleton P. P., 1983, ApJ 268, 368
  \bibitem{} Han Z., Podsiadlowski P., Eggleton P. P., 1994, MNRAS 270, 121
  \bibitem{} Han Z., Podsiadlowski P., Eggleton P. P., 1995, MNRAS 272, 800
  \bibitem{} Iben Jr. I., Livio M., 1993, PASP 105, 1373
  \bibitem{} King A. R., Begelman M. C., 1999, ApJ 519, 169
  \bibitem{} Nieuwenhuijzen H., de Jager C., 1990, A\&A 231, 134
  \bibitem{} Ostriker J., 1976, in: Structure and Evolution in Close Binary
             Systems, eds: P.~Eggleton, S.~Mitton, J.~Whelan, Proc. IAU 
             Symp. 73, Reidel, Dordrecht, p. 206
  \bibitem{} Paczynski B., 1976, in: Structure and Evolution of Close Binary 
             Systems, eds: P.~Eggleton, S.~Mitton, J.~Whelan, Proc. IAU 
             Symp. 73, Reidel, Dordrecht, p. 75
  \bibitem{} Pols O. R., Schr\"{o}der K.-P., Hurley J. R., Tout C. A., 
             Eggleton P. P. , 1998, MNRAS 298, 525
  \bibitem{} Pols O. R., Tout C. A., Eggleton P. P., Han Z., 1995, 
             MNRAS 274, 964
  \bibitem{} Tauris T. M., 1996, A\&A 315, 453
  \bibitem{} Tauris T. M., van den Heuvel E. P. J., Savonije G. J., 2000, 
             ApJ 530, L93
  \bibitem{} van den Heuvel E. P. J., 1994, A\&A 291, L39
  \bibitem{} Webbink R. F., 1984, ApJ 277, 355
\end{thebibliography}
\end{document}